\documentclass{article}

\usepackage[utf8]{inputenc}
\usepackage{latexsym}
\usepackage{amsmath,amssymb,amsfonts}

\usepackage{graphicx}

\newtheorem{definepart}{\sc Definition}[section]
\newtheorem{lemmapart}{\sc Lemma}[section]
\newtheorem{theorempart}{\sc Theorem}
\newtheorem{corollarypart}[theorempart]{\sc Corollary}
\newtheorem{assumepart}{\sc Assumption}[section]

\newenvironment{lemma}{\begin{lemmapart}\sl}{\end{lemmapart}}
\newenvironment{corollary}{\begin{corollarypart}\sl}{\end{corollarypart}}
\newenvironment{theorem}{\begin{theorempart}\sl}{\end{theorempart}}

\newenvironment{proof}{\noindent{\sc Proof:}}{\nobreak\hfill\nobreak$\Box$\par\medskip}

\makeatletter
\date{}
\title{Online Two-Dimensional Vector Packing with Advice}

\author{
Bengt J.~Nilsson\thanks{Malmö University, Sweden.} % email:~{\tt bengt.nilsson.TS@mah.se}
\and
Gordana Vujovic\thanks{University of Ljubljana, Slovenia.}% email:~{\tt }}
}
\makeatother

\newcommand{\figdir}{}
\newcommand{\setfigdir}[1]{\renewcommand{\figdir}{#1}}

\def\makespecial#1=#2.pdf{\special{pdffile=\figdir #2.pdf}}

\let\includegraphix=\includegraphics
\renewcommand{\includegraphics}[1]{\includegraphix{\figdir #1}}
\def\color[rgb]#1{}

\def\SetFigFont#1#2#3#4#5{}

\newlength{\capw}
\long\def\@caption#1[#2]#3{\par\addcontentsline{\csname
  ext@#1\endcsname}{#1}{\protect\numberline{\csname 
  the#1\endcsname}{\ignorespaces #2}}\begingroup
    \@parboxrestore
    \small
    \@makecaption{\csname fnum@#1\endcsname}{\ignorespaces #3}\par
\endgroup}

\newcommand{\parcaption}[2]{\parbox[t]{#1}{\caption{#2}}}

\setfigdir{Figures/}
\newcommand{\defeq}{\stackrel{{\rm def}}{=}}
\newcommand{\ld}[1]{\ensuremath{{\rm ld}(#1)}}
\newcommand{\OPT}{\ensuremath{{\rm OPT}}}

\newcommand{\sdk}[1]{\ensuremath{{\rm sc}_k(#1)}}

\newcommand{\TT}{\ensuremath{{\cal T}}}

\newcommand{\remove}[1]{\relax}

\begin{document}
\maketitle

\begin{abstract}
We consider the online two-dimensional vector packing problem, showing a lower bound of $11/5$ on the competitive ratio of any {\sc AnyFit} strategy for the problem. We provide strategies with competitive ratio $\max\!\left\{2,6\big/\big(1+3\tan(\pi/4-\gamma/2)\big)+\epsilon\right\}$ and logarithmic advice, for any instance where all the input vectors are restricted to have angles in the range $[\pi/4-\gamma/2,\pi/4+\gamma/2]$, for $0\leq\gamma<\pi/3$ and $\max\left\{5/2,4\big/\big(1+2\tan(\pi/4-\gamma/2)\big)+\epsilon\right\}$ and logarithmic advice, for any instance where all the input vectors are restricted to have angles in the range $[\pi/4-\gamma/2,\pi/4+\gamma/2]$, for $0\leq\gamma\leq\pi/3$. In addition, we give a $5/2$-competitive strategy also using logarithmic advice for the unrestricted vectors case. These results should be contrasted to the currently best competitive strategy, {\sc FirstFit}, having competitive ratio~$27/10$.
%\keywords{Bin Packing  \and Vector Packing \and Advice Complexity.}
\end{abstract}

\section{Introduction}\label{sec:intro}
Arguably, the problem of packing items into bins is among the most well-studied in computer science. It asks for the ``minimum number of unit sized bins required to pack a set of items, each of at most unit size,'' and has been shown to be NP-hard~\cite{garey1979computers:binpacking} but admits a PTAS~\cite{fer1981bin:offline}. It is common to view the bin packing problem through the lens of {\em online computation}, where the items are delivered one by one and each item has to be packed, either in an existing bin or a new bin, before the next item arrives. The quality of online strategies is measured by their {\em competitive ratio}, the worst case asymptotic or absolute ratio between the quality of the strategy's solution and that of an optimal one. Currently, the best strategy for online bin packing has asymptotic competitive ratio~$1.5815\!\ldots$~\cite{van2016beating} and it has been shown that no strategy can have asymptotic competitive ratio better than~$248/161=1.54037\!\ldots$~\cite{balogh2012new}. For the absolute competitive ratio the tight bound of $5/3$ has been proved~\cite{balogh2015absolute}.

The {\em vector packing problem\/} is a natural generalization of bin packing, where each item is a vector from $[0,1]^D$ and items are to be packed in $D$-dimensional unit cubes. Approximation algorithms with linear dependency on $D$ exist~\cite{chekuri2004multidimensional,fer1981bin:offline}. The online version of vector packing is not as well understood, the {\sc FirstFit} strategy has been shown to have competitive ratio $D+7/10$ even for the more general {\em resource constrained scheduling\/} problem~\cite{garey1976resource}. Azar~{\em et~al}.~\cite{azar2016packing:vp} claim that no online strategy for $D$-dimensional vector packing can have competitive ratio better than $D$ but offer no proof of this. They show however, that if all the vectors have $L_{\infty}$-norm at most $\epsilon^2$, there is a $(4/3+\epsilon)$-competitive algorithm for online two-dimensional vector packing. They also provide a $4/3$ lower bound for arbitrarily small vectors. In general, Galambos~{\em et~al}.~\cite{galambos1993lower:vp} provide a succinct lower bound for online $D$-dimensional vector packing that increases with $D$ but remains below~$2$ for all $D$. Their result implicitly gives a lower bound of $1.80288\!\ldots$, for $D=2$ which is currently the best known. Recently, almost ${\rm \Omega}(D)$ asymptotic lower bounds have been established for online $D$-dimensional vector packing for sufficiently large $D$~\cite{azar2013d-dim:lowerbound,balogh2020truly,bansal2020asymptotic}.

In many cases, the online framework is too restrictive in that it allows an all-powerful adversary to construct the input sequence in the worst possible way for the strategy. To alleviate this, {\em the advice complexity model\/} was introduced and has successfully yielded improved competitive ratios for bin packing and similar problems; see~\cite{angelopoulos2018online,Boyaretal:advice,bockenhauer2009advice:advice,Renetal:advice} for a selection of results. In this model, an oracle that knows both the online strategy and the input sequence provides the strategy with some prearranged information, {\em the advice}, about the input sequence, thus enabling the strategy to achieve improved competitive ratio.

%\vspace*{-2ex}
\subsection{Our Results}

We consider the general online two-dimensional vector packing problem. We begin by showing a lower bound of $11/5$ for the competitive ratio of any {\sc AnyFit} strategy~\cite{johnson1974fast:AF}  for the problem. In Section~\ref{sec:slaarv}, we provide a strategies with competitive ratio $\max\!\left\{2,6/\big(1+3\tan(\pi/4-\gamma/2)\big)+\epsilon\right\}$ and logarithmic advice, for any instance where all the input vectors are restricted to have angles in the range $[\pi/4-\gamma/2,\pi/4+\gamma/2]$, for $0\leq\gamma<\pi/3$ and $\max\left\{5/2,4\big/\big(1+2\tan(\pi/4-\gamma/2)\big)+\epsilon\right\}$ and logarithmic advice, for any instance where all the input vectors are restricted to have angles in the range $[\pi/4-\gamma/2,\pi/4+\gamma/2]$, for $0\leq\gamma\leq\pi/3$. In Section~\ref{sec:aslinagv}, we give a $5/2$-competitive strategy also using logarithmic advice for the unrestricted vectors case. These results should be contrasted to the currently best competitive strategy, {\sc FirstFit}, where an item is placed in the first bin where it fits, having competitive ratio~$27/10$~\cite{garey1976resource}.

\section{Preliminaries}\label{sec:prel}

We will use two norms in the sequel. For a two-dimensional vector $v$, the $L_1$-norm of $v$ is $\|v\|_1\defeq|v_x|+|v_y|$ and the $L_{\infty}$-norm (or max-norm) of $v$ is $\|v\|_{\infty}\defeq\max\{|v_x|,|v_y|\}$, where $v_x$ and $v_y$ are the $x$- and $y$-coordinates of~$v$, respectively. 

The {\em online vector packing problem\/} we consider is, given an input sequence $\sigma=(v_1,v_2,\ldots)$, of two-dimensional vectors $v_i\in[0,1]^2$, find the minimum number of unit sized square bins that can be packed online with vectors from the input sequence $\sigma$. From this problem definition we have that $0\leq v_x\leq1$ and $0\leq v_y\leq1$, i.e., all coordinates are non-negative.

A {\em packing\/} is simply a partitioning of the vectors into bins $B_1,B_2,\ldots$ such that for each bin $B_j$%\vspace*{-2ex}
\begin{align}\label{eqn:feasible}
\Big\|\sum_{v\in B_j}v\Big\|_{\infty}\leq1.
\end{align}%\vspace*{-2ex}

In the {\em online\/} packing variant, the vectors are released from the sequence one by one and a strategy that solves the packing problem must irrevocably assign a vector to a bin before the next vector arrives. The assignment is either to an already open bin, maintaining the feasibility requirement in Inequality~(\ref{eqn:feasible}), or the strategy must open a new bin and assign the vector to this bin. 

We measure the quality of an online strategy by its {\em competitive ratio}, the worst case bound $R$ such that $|A(\sigma)|\leq R\cdot|\OPT(\sigma)|+C$, for every possible input sequence $\sigma$, where $A(\sigma)$ is the solution produced by the strategy $A$ on $\sigma$, $\OPT(\sigma)$ is a solution on $\sigma$ for which $|\OPT(\sigma)|$ is minimal, and $C$ is some constant. 

In certain situations, the complete lack of information about future input is too restrictive. In a sense, the online strategy plays a game against an all-powerful adversary who can construct the input sequence in the worst possible manner. To alleviate the adversary's advantage, we consider the following {\em advice-on-tape\/} model~\cite{bockenhauer2009advice:advice}. An {\em oracle\/} has knowledge about both the strategy and the full input sequence from the adversary, it writes information on an {\em advice tape\/} of unbounded length. The strategy can read bits from the advice tape at any time, before or while the requests are released by the adversary. The {\em advice complexity\/} is the number of bits read from the advice tape by the strategy. Since the length of the advice bit string is not explicitly given, the oracle is unable to encode information into the length of the string, thereby requiring some mechanism to infer how many bits of advice the strategy should read at each step. This can be done with a self-delimiting encoding that extends the length of the bit string only by an additive lower order term~\cite{Boyaretal:advice}. 
A bit string $s$ is encoded as $e(s)=u(s)\circ b(s)\circ s$ ($\circ$ denotes concatenation), where $b(s)$ is a binary encoding of the length of the string $s$ and $u(s)$ consists of $\big|b(s)\big|$ ones followed by a single zero, thus indicating how many bits the strategy needs to read in order to obtain the length of the string $s$. The encoding has length at most $\big|e(s)\big|=|s|+2\lceil\log(|s|+1)\rceil+1$.
We henceforth assume that all advice information is encoded in this way.

We define the {\em load\/} of a bin $B$ to be the sum of the $L_1$-norms of the included vectors, i.e.,%\vspace*{-2ex}
\begin{align}
\ld{B}&\defeq\sum_{v\in B}\|v\|_1 = \sum_{v\in B}|v_x|+|v_y| = \sum_{v\in B}v_x+v_y,
\end{align}%\vspace*{-2ex}\\
since all coordinates are non-negative. The load for the whole request sequence $\sigma$ is
\begin{align}
\ld{\sigma}\defeq\sum_{v\in\sigma}\|v\|_1=\sum_{v\in\sigma}v_x+v_y. 
\end{align}
Since the maximum load in a bin is 2, we immediately have that%\vspace*{-2ex}
\begin{align}\label{eqn:loadbound}
|\OPT(\sigma)|\geq\lceil\ld{\sigma}/2\rceil.
\end{align}%\vspace*{-6.0ex}

\section{A Lower Bound for Two-Dimensional {A{\small NY}F{\small IT}} Strategies}\label{sec:albtdff}

Currently, the best lower bound on the competitive ratio for two-dimensional vector packing is $R\geq1.80288\!\ldots$, implicit from the construction by Galambos~{\em et al.}~\cite{galambos1993lower:vp}. We show here a lower bound for two-dimensional vector packing valid for the class of {\sc AnyFit} strategies. An online strategy $A$ is an {\sc AnyFit} strategy, if $A$ does not open a new bin unless the current item released from the input sequence does not fit in any already opened bin~\cite{johnson1974fast:AF}. 

Let $\sigma_1=(p_i)_{i=1}^{n}$, $0<p_i\leq1$, be an instance of the one-dimensional online bin packing problem for which {\sc AnyFit} strategy $A$ has competitive ratio at least $|A(\sigma_1)|\geq
\lfloor17|\OPT(\sigma_1)|/10\rfloor$, where $\OPT(\sigma_1)$ is an optimal solution. Such sequences $\sigma_1$ exist of arbitrary length; see Dósa and Sgall~\cite{dosa2013FF} and Johnson~\cite{johnson1974fast:AF}, and we chose $\sigma_1$ so that $|\OPT(\sigma_1)|$ is a multiple of~10.

To construct our lower bound for the two-dimensional case, let $p_{\min}=\min\{p_1,\ldots,p_n\}$, choose $0<\delta<p_{\min}/2$, and construct a two-dimensional instance $\sigma_2$ as follows. The sequence $\sigma_2$ has a {\em prefix\/} consisting of $|\OPT(\sigma_1)|$ copies of the vector $(0,1/2)$, followed by a {\em suffix}, the sequence $(p_1,\delta), (p_2,\delta),\ldots, (p_n,\delta)$. An optimal packing $\OPT(\sigma_2)$ has the same size as $\OPT(\sigma_1)$, since each bin in the optimal solution packs the vectors in the suffix optimally according to the $x$-coordinate and since we chose $\delta<p_{\min}/2$, the space used on the $y$-coordinate in each bin is $<1/2$, so one of the $(0,1/2)$ vectors can be placed in each bin. Thus, $|\OPT(\sigma_2)|=|\OPT(\sigma_1)|$.

The {\sc AnyFit} strategy $A$, given the vectors in $\sigma_2$, will pack the prefix %$|\OPT(\sigma_1)|$
vectors $(0,1/2)$ pairwise into $\lfloor|\OPT(\sigma_1)|/2\rfloor$ bins that are full with respect to the $y$-coordinate. No vector in the suffix can be packed in any of these bins as they all have positive $y$-coordinate. The suffix is therefore packed by $A$ into at least $\lfloor17|\OPT(\sigma_1)|/10\rfloor$ bins, giving us a total of at least $\lfloor|\OPT(\sigma_1)|/2\rfloor+\lfloor17|\OPT(\sigma_1)|/10\rfloor
=
|\OPT(\sigma_1)|/2+17|\OPT(\sigma_1)|/10
=
11|\OPT(\sigma_1)|/5$ bins, since $|\OPT(\sigma_1)|$ is a multiple of~10. 
We state this as a theorem.%\vspace*{-1ex}
\begin{theorem}\label{thm:lowerbound}
Every {\sc AnyFit} strategy $A$ has a lower bound for two-dimensional vector packing of%\vspace*{-1ex}
$$
\big|A(\sigma)\big|\geq
\frac{11}{5}\big|\OPT(\sigma)\big|,%\vspace*{-1ex}
$$
for some input sequence~$\sigma$.
\end{theorem}

The approach can easily be generalized to $D$ dimensions, achieving $$\big|A_D(\sigma)\big|\geq(D/2+6/5)\big|\OPT(\sigma)\big|,$$
for every {\sc AnyFit} strategy~$A_D$.

%\vspace*{-3ex}

\section{Strategies with Logarithmic Advice for Angle Restricted Vectors}\label{sec:slaarv}
\subsection{A First Strategy}\label{sec:afs}

We assume in this case that each vector $v$ in the input sequence $\sigma$ has angle%\vspace*{-1ex}
\begin{align}
\arg v & \in[\pi/4-\gamma/2,\pi/4+\gamma/2],
\end{align}
%\vspace*{-4ex}\\
for a given extremal angle $\gamma$; see Figure~\ref{limitedpartitionLNCS}. This set of angles forms a cone issuing from the origin towards the point $(1,1)$. Let $d=1+\tan(\pi/4-\gamma/2)$, i.e., the abscissa of the line passing through the intersection points of the horizontal and vertical lines through $(1,1)$. 
%This means that the competitive ratio of our algorithm will depend on~$\gamma$. 
\begin{figure*}
                %\vspace*{-2.0ex}
			    \begin{center} \small
			    \input{\figdirlimitedpartitionLNCS.pdf_t}
			    \parcaption{\capw}{\label{limitedpartitionLNCS}The partitioning of vectors into five groups.}
			    \end{center}
			    %\vspace*{-3.0ex}
			    \end{figure*}

We follow the exposition in the proof of Theorem~4 in~\cite{Boyaretal:advice} for the one-dimensional case %but instead of partitioning the vectors into four groups, we partition them into five groups. A vector $v$ is
and partition the vectors into five groups. A vector $v$ is
\begin{description}
\item 
Tiny: if $\|v\|_1\leq a$, ($a<1$ is some constant to be established later)

\item
Small: if $a < \|v\|_1\leq d/2$, %($d>2a$ is some constant to be established later) !!!!already presented

\item
Medium: if $d/2 < \|v\|_1\leq 1$, 

\item
Large: if $1 < \|v\|_1\leq b$, and ($1<b<2$ is some constant to be established later)

\item
Huge: if $b < \|v\|_1\leq 2$;
\end{description}
see Figure~\ref{limitedpartitionLNCS}.
To ensure that no small or medium vector can be packed together with a huge vector in a bin % and no five small vectors can be packed together in a bin
and no three small or medium vectors can be packed together in a bin, we enforce that $a+b\geq2$ and $3a\geq2$, giving us $a\geq2/3$ and $b\geq4/3$. Furthermore, $a<d/2$ implies that $d>4/3$, whereby $\gamma<\pi/3$. Furthermore, the fact that $d\geq4/3$, ensures that no three medium items can be packed together in a bin.

%We note that both the cases $b\leq d$ and $b\geq d$ are possible.

To bound the number of advice bits used, we fix a positive integer parameter~$k$.
Each region of large, medium, and small vectors, respectively, is partitioned into $k$ diagonal strips as $a+(i-1)(d/2-a)/k<\|v\|_1\leq a+i(d/2-a)/k$, for $1\leq i\leq k$ of the small vectors, $d/2+(i-1)(1-d/2)/k<\|v\|_1\leq d/2+i(1-d/2)/k$, for $1\leq i\leq k$ of the medium vectors, and $1+(i-1)(b-1)/k<\|v\|_1\leq 1+i(b-1)/k$, for $1\leq i\leq k$ of the large vectors.

The advice that the strategy obtains is the number of large, medium, and small vectors, respectively in each of the $k$ strips, thus $O(k\log n)$ bits of advice in total.

Our strategy $A_{\gamma}$ reads the $3k$ values $L_1,\ldots,L_k$, $M_1,\ldots,M_k$, and $S_1,\ldots,S_k$ corresponding to the number of vectors in each strip, lets $L=\sum_{i=1}^k L_i$, $M=\sum_{i=1}^k M_i$, $S=\sum_{i=1}^k S_i$, and opens $L+M+\lceil S/2\rceil$ bins, henceforth 
denoted large, medium and small {\em critical bins}. It reserves space $1+i(b-1)/k$ for each of $L_i$ large critical bins and $d/2+i(1-d/2)/k$ for each of $M_i$ medium critical bins, $1\leq i\leq k$. We say that a bin has a {\em virtual load\/} of $1+i(b-1)/k$ and $d/2+i(1-d/2)/k$, respectively. 
For the small critical bins, we reserve space matching the sum of the upper limit from the lowest non-empty small strip with the upper limit from the highest non-empty  small strip, for a pair of vectors that can be matched together, thus reducing the number of vectors in those strips by one each. If the two strips are $a+(i-1)(d/2-a)/k<x+y\leq a+i(d/2-a)/k$ and $a+(j-1)(d/2-a)/k<x+y\leq a+j(d/2-a)/k$ for $i\leq j$, we reserve the virtual load of $2a+(i+j)(d/2-a)/k$ to the bin. We repeat the process until at most a single small vector remains for which we reserve the virtual load of $a+i(d/2-a)/k$, if the vector is in the $i^{\rm th}$ strip. The reserved space in a bin is used to pack one large, one medium or two small vectors in the bin when the vector arrives. 
$A_{\gamma}$ then serves each vector $v$ in the sequence $\sigma$ in order
as follows:
%\vspace*{-1.5ex}
\begin{description}
\item 
if $v$ is huge, open a new bin and place $v$ in this bin,

\item
if $v$ is large and lies in strip $i$, place $v$ in the reserved space of the first unused large critical bin with reserved space $1+i(b-1)/k$, reduce the virtual load of the bin to the actual load (by the amount $1+i(b-1)/k - \|v\|_1$),

\item
if $v$ is medium and lies in strip $i$, place $v$ in the reserved space of the first unused medium critical bin with reserved space $d/2+i(1-d/2)/k$, reduce the virtual load of the bin to the actual load (by the amount $d/2+i(1-d/2)/k - \|v\|_1$),

\item
if $v$ is small and lies in strip $i$, place $v$ in the reserved space of the first small critical bin that contains at most one small vector and has unused reserved space $a+i(d/2-a)/k$, reduce the virtual load by $a+i(d/2-a)/k - \|v\|_1$ and, if the bin now contains two small vectors, update the actual load.

\item
if $v$ is tiny, use the {\sc FirstFit} strategy and place the vector in the first open bin where it fits based on virtual load (add the current vector to the virtual load, obtaining $\|v\|_1+x+y$) and if this is not possible, open a new bin and place $v$ in this bin.
\end{description}

\begin{theorem}\label{thm:limited-log2}
For any angle $0\leq\gamma<\pi/3$ and $\epsilon>0$, the strategy $A_{\gamma}$ receives $O(\frac{1}{\epsilon}\log n)$ bits of advice and has cost $c(\gamma,\epsilon)|\OPT(\sigma)| + 1$ for serving any sequence $\sigma$ of size~$n$, where%\vspace*{-2ex}
$$
c(\gamma,\epsilon)=\max\left\{2,\frac{6}{1+3\tan(\pi/4-\gamma/2)}+\epsilon\right\}.%\vspace*{-1ex}
$$
\end{theorem}
\begin{proof}
Assume first that our strategy uses $|A_{\gamma}(\sigma)|=H+L+M+\lceil S/2\rceil\leq H+L+M+S/2+1$ bins, i.e., there is no bin that only contains tiny vectors. The optimal strategy must use at least $|\OPT(\sigma)|\geq H+L$ bins since no two huge or large vectors can be placed in the same bin. If $L\geq M+S$, then the number of bins that our strategy uses is%\vspace*{-1ex}
\begin{align}
|A_{\gamma}(\sigma)| 
&
\leq
H+L+M+S/2+1
\leq
H+L+M+S+1
\leq
H+2L+1
\nonumber\\&
\leq 
2H+2L+1
\leq
2|\OPT(\sigma)|+1.
\end{align}

%\vspace*{-1ex}%
On the other hand, if $L<M+S$, then the optimal strategy can only place one medium or small vector together with a large one in a bin and the remaining medium and small vectors can at best be packed together two and two, the optimal strategy must therefore use at least $|\OPT(\sigma)|\geq H+L+(M+S-L)/2=H+L/2+M/2+S/2$ bins. The number of bins that our strategy then uses is%\vspace*{-1ex}
\begin{align}
|A_{\gamma}(\sigma)| &
\leq
H+L+M+S/2+1
\leq
2H+L+M+S+1
\nonumber\\&
=
2(H+L/2+M/2+S/2)+1
\leq
2|\OPT(\sigma)|+1.
\end{align}

%\vspace*{-1ex}%
Assume now that our strategy constructs at least one bin with only tiny vectors in it. Any of these vectors did not fit in any of the critical bins initially opened, thus the virtual load of each critical bin is at least $d-a$. Since the difference between virtual load and actual load is at most $1/k$ (we know the number of vectors in each strip), the actual load is at least $d-a-1/k$. Since the bins with huge vectors also have a load of this magnitude and we can have at most one bin with tiny vectors having less than this load, all but one bin have load at least $d-a-1/k$. We have, if our strategy uses $|A_{\gamma}(\sigma)|$ bins, that%\vspace*{-1ex}
\begin{align}\label{eqn:binbound}
\ld{\sigma} = \sum_{i=1}^{|A_{\gamma}(\sigma)|} \ld{B_i} \geq (|A_{\gamma}(\sigma)|-1)(d-a-1/k),
\end{align}%\vspace*{-3ex}\\
so our strategy uses%\vspace*{-1ex}
\begin{align}
|A_{\gamma}(\sigma)|
&\leq
\frac{\ld{\sigma}}{d-a-1/k} + 1
\leq
\frac{2|\OPT(\sigma)|}{d-a-1/k} + 1
%\nonumber\\&
\leq
\frac{2|\OPT(\sigma)|}{\tan(\pi/4-\gamma/2)+1/3-1/k} + 1.
\end{align}%\vspace*{-2ex}\\
bins, by Inequality~(\ref{eqn:loadbound}) and since $d=1+\tan(\pi/4-\gamma/2)$ and $a=2/3$.
%, the number of bins used is\vspace*{-1ex}
%\begin{align}
%|A_{\gamma}(\sigma)|\leq \frac{2|\OPT(\sigma)|}{\tan(\pi/4-\gamma/2)+1/3-1/k} + 1.
%\end{align}\vspace*{-2ex}\\

Choosing $\epsilon=8/k$, for $k\geq8$, we have the competitive ratio as claimed, since $\tan(\pi/4-\gamma/2)>1/4$ for every~$\gamma\in[0,\pi/3[$.

%The advice that the strategy receives consists of $3k$ values, % denoting
The strategy receives $3k$ values as advice,
the number of vectors in each strip. Each value is encoded with at most $\lceil\log(n+1)\rceil$ bits, where $n=|\sigma|$. %Consequently, 
Hence, the number of advice bits is at most $3k\big(\lceil\log(n+1)\rceil\big)\in O\big(\frac{1}{\epsilon}\log n\big)$.
%\hfill$\Box$
\end{proof}

\subsection{A Second Strategy}\label{sec:ass}
If we modify the strategy $A_{\gamma}$ in such a way that no huge vector can be packed together with a small vector in a bin, no two large vectors can be packed in a bin, no three medium vectors can be packed in a bin, and no four small vectors can be packed in a bin, we get $a+b\geq2$, $4a\geq2$ and $3d/2\geq2$, giving us $a\geq1/2$, $b\geq3/2$, and $d/2\geq2/3$, again requiring that~$\gamma\leq\pi/3$.
\begin{theorem}\label{thm:limited-log2.5}
For any angle $0\leq\gamma\leq\pi/3$ and $\epsilon>0$, the modified strategy $A'_{\gamma}$ described receives $O(\frac{1}{\epsilon}\log n)$ bits of advice and has cost $c(\gamma,\epsilon)|\OPT(\sigma)| + 1$ for serving any sequence $\sigma$ of size~$n$, where $$c(\gamma,\epsilon)=\max\left\{\frac{5}{2},\frac{4}{1+2\tan(\pi/4-\gamma/2)}+\epsilon\right\}.$$
\end{theorem}
\begin{proof}
As in the previous proof, assume first that our modified strategy uses $|A'_{\gamma}(\sigma)|= H+L+M+\lceil S/2\rceil\leq H+L+M+S/2+1$ bins, i.e., there is no bin that only contains tiny vectors.
If we fix an optimal solution $\OPT(\sigma)$, this solution must contain at least $|\OPT(\sigma)|\geq H+L$ bins since no two huge or large vectors can be placed in the same bin. If $3L/2\geq M+S/2$, then our strategy uses 
\begin{align}
|A'_{\gamma}(\sigma)| & \leq H+L+M+S/2+1
\leq
H+5L/2 +1
\leq 
5(H+L)/2+1
%\nonumber\\
%&
\leq
5|\OPT(\sigma)|/2+1
\end{align}
bins.

On the other hand, if $3L/2<M+S/2$, then we consider two further cases.
If $S\leq(M-L)/2$, then the best case for $\OPT(\sigma)$ is that each bin with a large vector also contains a medium vector and the remaining medium vectors are packed two by two together with a small vector. Hence,
\begin{align}
|\OPT(\sigma)| & \geq H+L + (M-L)/2 = H+L/2 +M/2.
\label{eqn:lb0}
\end{align}
So, our strategy uses 
\begin{align}
|A'_{\gamma}(\sigma)| & \leq H+L+M+S/2+1
\leq
H+L+M+\frac{(M-L)/2}{2}+1
\nonumber\\
&
\leq 
5(H+L/2+M/2)/2+1
\leq
5|\OPT(\sigma)|/2+1
\end{align}
bins.

Now, if $S>(M-L)/2$, then the best case for $\OPT(\sigma)$  is that each bin with a large vector also contains a medium or a small vector (if needed), each bin with two medium vectors also contain a small vector (if needed), each bin with one medium vector contains two small vectors (if needed), and that the remaining small vectors can be packed three and three (if needed). This means that if each of the small vectors is packed together with a large vector in a bin, then 
\begin{align}
|\OPT(\sigma)| & \geq H+L + (M-(L-S))/2 = H+L/2 +M/2+S/2
\label{eqn:lb1}
\end{align}
and, otherwise not all the small vectors are packed together with large vectors, giving us
\begin{align}
|\OPT(\sigma)| \geq H+L + (M+S-L)/3 = H+2L/3 +M/3+S/3.
\label{eqn:lb2}
\end{align}
The right hand expression of inequality~(\ref{eqn:lb1}) is smaller than the right hand expression of inequality~(\ref{eqn:lb2}) when $S+M<L$, contradicting that $L\leq3L/2<M+S/2\leq M+S$. 
Thus, inequality~(\ref{eqn:lb2}) is a lower bound when~$3L/2<M+S/2$.

Since $3L/2<M+S/2$ and $S>(M-L)/2$, then we have
\begin{align}
|A'_{\gamma}(\sigma)| & \leq H+L+M+S/2+1
<
H+L+5M/6+S/2+1 + (L+2S)/6
\nonumber\\ & 
=
H+7L/6+5M/6+5S/6+1
\leq
5(H+2L/3+M/3+S/3)/2+1
\nonumber\\ & 
\leq
5|\OPT(\sigma)|/2+1
\end{align}
bins, where the second inequality uses $M< L + 2S$.

The case that our new strategy constructs a bin with only tiny vectors in it is handled exactly as in the proof of Theorem~\ref{thm:limited-log2} giving us
\begin{align}
|A'_{\gamma}(\sigma)| & 
\leq 
\frac{2|\OPT(\sigma)|}{d-a-1/k} + 1
\leq 
\frac{2|\OPT(\sigma)|}{\tan(\pi/4-\gamma/2)+1/2-1/k} + 1,
\end{align}
since $d=1+\tan(\pi/4-\gamma/2)$ and~$a=1/2$.
Choosing $k\geq8/\epsilon$, we have the lemma as claimed, since $\tan(\pi/4-\gamma/2)$ is minimized for~$\gamma=\pi/3$.

We also omit the argument for the size of the advice, since it follows exactly the same lines as in the proof of Theorem~\ref{thm:limited-log2}.
\end{proof}%
\begin{figure*}
                %\vspace*{-2.0ex}
			    \begin{center} \small
			    \input{\figdirlimitedplot.pdf_t}
			    \parcaption{\capw}{\label{limitedplot}Plots for worst case competitive ratios of solution $A_{\gamma}(\sigma)$ (red) and solution $A'_{\gamma}(\sigma)$ (green), for $\gamma\in[0,\pi/3]$.}
			    \end{center}
			    %\vspace*{-3.0ex}
			    \end{figure*}%

We refer to Figure~\ref{limitedplot} for an illustration of the worst case competitive ratio for Theorems~\ref{thm:limited-log2} and~\ref{thm:limited-log2.5}. As we can see, Theorem~\ref{thm:limited-log2.5} guarantees a smaller competitive ratio for larger values of $\gamma$. The break-even point is for $\gamma = \pi/2 - 2\arctan(7/15) \approx 0.69754 \approx 0.2220\pi$. Furthermore, Theorem~\ref{thm:limited-log2.5} guarantees a competitive ratio of at most $2/(\tan(\pi/12) + 1/2) + \epsilon \approx 2.6043 + \epsilon$ improving on {\sc FirstFit} for the full range of applicable maximum angles values~$\gamma$.

\section{A Strategy with Logarithmic Advice for General 2D-Vectors}\label{sec:aslinagv}

We generalize the approach by Fernandez de la Vega and Lueker~\cite{fer1981bin:offline} for the one-dimensional case and let $k$ be a positive integer. Subdivide the unit square (representing the bins) by a $(k+1)\times(k+1)$ grid with intersection points at $(i/k,j/k)$, for $0\leq i,j\leq k$. The region $](i-1)/k,i/k]\times](j-1)/k,j/k]$, for $1\leq i,j\leq k$, is called the {\em $(i,j)$-box}. (Except for the special case when $i=1$ or $j=1$, then the $(1,1)$-box is the region $[0,1/k]\times[0,1/k]$, the $(1,j)$-box is the region $[0,1/k]\times](j-1)/k,j/k]$, and the $(i,1)$-box is the region $](i-1)/k,i/k]\times[0,1/k]$.) When the specific coordinates of an $(i,j)$-box are unimportant, we will simply refer to it as a {\em box}. A vector $v$ such that $(i-1)/k< v_x\leq i/k$ and $(j-1)/k< v_y\leq j/k$ is said to {\em lie in\/} or {\em be contained in\/} the $(i,j)$-box. (Again, the special case when $v_x=0$ or $v_y=0$, the vector lies in the $(1,j)$-box or the $(i,1)$-box respectively.) We say that a vector $v$ with $v_x\leq40/k$ and $v_y\leq40/k$ is {\em short}. All other vectors are {\em long\/}; see Figure~\ref{squarepartition}(a).%
\begin{figure*}
                %\vspace*{-2.0ex}
			    \begin{center} \small
			    \input{\figdirsquarepartition.pdf_t}
			    \parcaption{\capw}{\label{squarepartition}(a) The partitioning of vectors into boxes ($k=10$ for illustration). Vectors in green region are short (not to scale for purpose of illustration), a long vector $v$ (blue) and $v$ $k$-scaled to \sdk{v} (red). (b) Illustrating the proof of Lemma~\protect\ref{lem:pack2.5}.}
			    \end{center}
			    %\vspace*{-3.0ex}
			    \end{figure*}

Let \sdk{v} denote the {\em $k$-scaled\/} vector $v$, where $\sdk{v}=(i/k,j/k)$, if $v$ lies in the $(i,j)$-box. $k$-Scaling the vectors in $\sigma$ reduces the types of vectors from possibly $|\sigma|$ to $k^2$. Disregarding the short vectors, the number of long vectors that can appear in a bin is at most $2\lfloor k/40\rfloor\leq k/20$, since we can fit at most $\lfloor k/40\rfloor$ long horizontal vectors and at most $\lfloor k/40\rfloor$ long vertical vectors in a bin.

Let $\sigma_L$ be the subsequence of long vectors in $\sigma$ and let $\sigma_S$ be the subsequence of short vectors in $\sigma$, both dependent on the parameter $k$. Let \sdk{\sigma_L} denote the $k$-scaled vectors in $\sigma_L$ and let $\OPT\big(\sdk{\sigma_L}\big)$ be an optimal solution of the $k$-scaled long vectors in~\sdk{\sigma_L}. %Each long vector \sdk{v} in \sdk{\sigma_L} is packed optimally in some bin that is of one out of the $O(k^{k/10})$ bin types and therefore has some binary description of size~$b\in O(k\log k)$. As $v$ is released on the input tape, if $v$ is short, the strategy reads $b$ zeros from the advice tape, and if $v$ is long, the strategy reads the $b$ bits representing the bin type containing \sdk{v}\ in~$\OPT\big(\sdk{\sigma_L}\big)$.

We let the advice given by the oracle be the number of long vectors in each box. (Note that information about short vectors is not provided.) Given the number of long vectors, $n_{i,j}$, in each box, $1\leq i,j\leq k$, a brute force algorithm can compute an optimal solution $\OPT\big(\sdk{\sigma_L}\big)$ in time polynomial in $|\sigma_L|\leq|\sigma|=n$; see the proof of Theorem~\ref{thm:log2.5}. We let our strategy $A_k$ perform this computation and open a critical bin corresponding to each bin in the solution $\OPT\big(\sdk{\sigma_L}\big)$. To each critical bin we reserve space corresponding to the sum of  the $L_1$-norms of the assumed $k$-scaled vectors placed in it and set the virtual load of the bin to be this value. As the strategy serves requests from the sequence $\sigma$, each long vector $v$ contained in an $(i,j)$-box is placed in the first bin that has remaining space for a $k$-scaled vector in an $(i,j)$-box. The virtual load of the bin is reduced by the difference $i/k+j/k-\|v\|_1$. Each short vector that arrives is placed according to the {\sc FirstFit} rule in the first bin where it fits according to the current virtual load and the virtual load is increased accordingly. If no such bin exists, the strategy opens a new bin and places the short vector there.

%Let $t$ be the bit string of size $b$ describing the bin type used in $\OPT\big(\sdk{\sigma_L}\big)$ to pack the vector \sdk{v}. The string $t$ is read from the advice tape as we read the description of $v$ from the input. Our strategy maintains two types of bins, {\em long\/} and {\em short\/} bins.  Long bins are used to pack the long vectors and short bins pack the short vectors as follows:
%
%if $v$ is long, our strategy simply tests if a long bin of type $t$ already exists and still has space for the allocated scaled vector \sdk{v}. If so, it places $v$ in this bin, otherwise it opens a long bin of type $t$ and places $v$ in this bin. If $v$ is short, it adds $v$ to the currently open short bin (the strategy keeps only one open short bin at any time) if it can be added, otherwise it closes the currently open short bin and opens a new empty short bin and places $v$ here.

\medskip
We first show a lower bound for our strategy $A_k$.
Assume that $k$ is odd and consider $2s$ copies of the vector $(1/2-\epsilon,\epsilon)$ and $s$ copies of the vector $(2\epsilon,1-2\epsilon)$, for sufficiently small $\epsilon<1/3k$. An optimal packing of these vectors uses $s$ bins but the $k$-scaled vectors become $2s$ copies of $(1/2+1/2k,1/k)$ and $s$ copies of $(1/k,1)$. No two scaled vectors fit together in a bin, hence an optimal packing of scaled vectors requires $3s$ bins, giving us a competitive ratio of at least~$3$.

\medskip
However, when $k$ is even, we can show that the competitive ratio of $A_k$ is $5|\OPT(\sigma)|/2+1$ by first proving the following lemma.%\vspace*{-1.0ex}%
\begin{lemma}\label{lem:pack2.5}
For $k\geq100$ and even, %\vspace*{-3ex}%
$
\big|\OPT\big(\sdk{\sigma_L}\big)\big| \leq \dfrac{5}{2}\big|\OPT(\sigma_L)\big| + 1.%\vspace*{-1ex}%
$
\end{lemma}%\vspace*{-2.5ex}%
\begin{proof}
We prove that for any bin $B$ packed with long vectors, the corresponding $k$-scaled vectors can be packed into at most two bins and one half bin. A {\em half bin\/} is a 2-dimensional bin of size $[0,1/2]\times[0,1/2]$. Let $Q\big(\sdk{\sigma_L}\big)$ be such a repacking of the bins in $\OPT(\sigma_L)$. This means that $Q\big(\sdk{\sigma_L}\big)$ consists of at most $2|\OPT(\sigma_L)|$ bins and $|\OPT(\sigma_L)|$ half bins. Of course, the vectors in any two half bins can be packed together into one bin, giving us a new repacking $R\big(\sdk{\sigma_L}\big)$ of size%\vspace*{-1.0ex}%
\begin{align}
\big|R\big(\sdk{\sigma_L}\big)\big| 
& \leq
2\big|\OPT(\sigma_L)\big| + \left\lceil\big|\OPT(\sigma_L)\big|/2\right\rceil
\leq
%\lceil5|\OPT(\sigma_L)|/2\rceil
%\leq
5\big|\OPT(\sigma_L)\big|/2+1
\end{align}%\vspace*{-3ex}\\%
bins. Since $R\big(\sdk{\sigma_L}\big)$ is a feasible packing of the long vectors, we have that $\big|\OPT\big(\sdk{\sigma_L}\big)\big|\leq\big|R\big(\sdk{\sigma_L}\big)\big|$ and the result as claimed.

Let $B$ be an arbitrary bin and assume that $B$ contains $m\geq1$ long vectors. Assume the vectors are ordered $v_1,\ldots,v_m$ by decreasing $L_1$-norms of their $k$-scaled corresponding vectors, i.e.,  $\|\sdk{v_1}\|_1\geq\|\sdk{v_2}\|_1\geq\cdots\geq\|\sdk{v_m}\|_1$. For ease of notation we use $v'_i\defeq\sdk{v_i}$, for $1\leq i\leq m$. Let $x_{i,j}$ and $y_{i,j}$ denote the sum of the $x$-coordinates and $y$-coordinates, respectively of $v'_i,\ldots,v'_j$, for $i\leq j$, according to our ordering. It is clear that $x_{i,j}\leq1+(j-i+1)/k$ and $y_{i,j}\leq1+(j-i+1)/k$, for any $1\leq i\leq j\leq m$.

By construction, any single vector $v'_i$ fits in one bin so we can assume that $m\geq2$. Also, if $m=2$, then the two vectors can be packed in two separate bins, immediately proving our result, hence we assume $m\geq3$. Consider the sequence $v'_1+v'_2+\cdots+v'_m$, starting at the origin in the bin; see Figure~\ref{squarepartition}(b). Fix $v'_a$ to be the first vector that intersects the exterior of the bin ($v'_a$ must exist, otherwise the whole sequence fits in the bin, immediately proving our claim) and assume without loss of generality that it intersects the vertical boundary of the bin. (The other case, where the sequence intersects the horizontal boundary is completely symmetric.) Obviously,~$a\geq2$.

Since the bin $B$ has $m$ long vectors and $m\leq k/20$, we have that $x_{1,m}\leq1+m/k\leq21/20$ and symmetrically $y_{1,m}\leq1+m/k\leq21/20$. Also, by our assumption that $v'_a$ is the first vector intersecting the exterior of the bin, we have $x_{a+1,m}\leq1/20$; see Figure~\ref{squarepartition}(b).
We make the following case analysis:
\begin{description}
\item[]if $x_{a,m}\leq1$ and $y_{a,m}\leq1$, 
then we can pack the vectors $v'_1,\ldots,v'_{a-1}$ in one bin and the vectors $v'_a,\ldots,v'_m$ in a second bin, satisfying our requirement.

\item[]if $x_{a,m}\leq1$ and $y_{a,m}>1$, we have two further cases:
\begin{description}
\item[]if there is a vector $v'_b$, $a<b\leq m$, such that $y_{a,b-1}\leq1$ and $y_{b,m}\leq1/2$, then we can pack $v'_1,\ldots,v'_{a-1}$ in one bin, $v'_a,\ldots,v'_{b-1}$ in a second bin, and $v'_b,\ldots,v'_{m}$ in a half bin.

%\item[]if $y_{m,m}>1/2$, then the $L_1$-norm $\|v'_m\|_1\geq y_{m,m}>1/2$, so each vector $v'_1,\ldots,v'_{m-1}$ must also have $L_1$-norm greater than $1/2$, thus $m\leq3$ in this case. If $m\leq2$, we can pack the vectors into at most two bins, satisfying our requirement. Assume therefore that $m=3$. Since $x_{1,3}\leq1+3/k$ and $y_{1,3}\leq1+3/k$, each of the three vectors can have one coordinate of size at least $1/2+1/k$ (since $k$ is even) making them impossible to pack into two bins and a half bin. However, this means that the three vectors $v_1$, $v_2$, and~$v_3$ each have a coordinate strictly larger than $1/2$ and they can therefore not be packed together in one bin, a contradiction. Hence, at least one of the three vectors $v'_1$, $v'_2$, and~$v'_3$ have both coordinates at most $1/2$ and we can therefore pack the three vectors into two bins and one half bin. 

\item[]if there is a vector $v'_b$, $a<b\leq m$, such that $y_{a,b-1}\leq1$, $y_{a,b}>1$, and $y_{b,m}>1/2$, then since $y_{a,m}\leq y_{1,m}\leq21/20$, we have that $y_{b+1,m}=y_{a,m}-y_{a,b}<21/20-1=1/20$, if the sequence $v'_{b+1},\ldots,v'_{m}$ exists. Thus, $y_{b,b}=y_{b,m}-y_{b+1,m}>1/2-1/20=9/20$, whether or not the sequence $v'_{b+1},\ldots,v'_{m}$ exists. Since the $L_1$-norm $\|v'_b\|_1\geq y_{b,b}>9/20$, each vector $v'_1,\ldots,v'_{b-1}$ must also have $L_1$-norm greater than $9/20$, thus $b\leq4$ in this case. Since $a\geq2$, it follows that~$b\in\{3,4\}$. Reorder $v'_1,\ldots,v'_{b-1}$ so that $x_{1,1}\geq\cdots\geq x_{b-1,b-1}$.

\medskip
If $b=3$, we have three cases.
\begin{description}
\item[]If $9/20<y_{3,3}\leq1/2$, then both $y_{1,1}<1+3/k-9/20=11/20+3/k<12/20=3/5$ and $y_{2,2}<1+3/k-9/20=11/20+3/k<12/20=3/5$. Since $x_{1,1}>1/2$ and $x_{2,2}>1/2$ is not possible, otherwise $v_1$ and $v_2$ would not fit together in one bin as they would both have $x$-coordinate greater than 1. We can pack $v'_1$ in a bin, $v'_2$ and $v'_4,\ldots,v'_m$ (if they exist) in a second bin, and $v'_3$ in a half bin.
\item[]If $1/2<y_{3,3}\leq19/20$, and again $x_{2,2}\leq1/2$ so we can pack $v'_1$ in a bin, $v'_2$ in a half bin, and $v'_3$ and $v'_4,\ldots,v'_m$ (if they exist) in a second bin.
\item[]If $19/20<y_{3,3}\leq1$, and again $x_{2,2}\leq1/2$. If $x_{1,1}>19/20$, then $\|v'_2\|_1<2+6/k-19/20-19/20=1/10+6/k<3/20$, a contradiction. Hence, $x_{1,1}\leq19/20$ and we can pack $v'_1$ and $v'_4,\ldots,v'_m$ (if they exist) in a bin, $v'_2$  in a half bin, and $v'_3$ in a second bin.
\end{description}

\medskip
If $b=4$, then since each vector $v'_1$,~$v'_2$,~$v'_3$, and~$v'_4$ has $L_1$-norm greater than $9/20$, each of them also has $L_1$-norm smaller than $2+8/k-3\cdot9/20=13/20+8/k$. Hence, we have $x_{1,1}<13/20+8/k<14/20=7/10$, $x_{2,2}<7/10$, and~$x_{3,3}<7/10$.  We have two cases.
\begin{description}
\item[]If $9/20<y_{4,4}\leq1/2$, then if $x_{1,1}\leq1/2$, then we can pack $v'_1$ and $v'_2$ in a bin, $v'_3$ and $v'_5,\ldots,v'_m$ (if they exist) in a second bin, and $v'_4$ in a half bin. If $x_{1,1}>1/2$ then, since $x_{2,3}=x_{1,3}-x_{1,1}<1+3/k-1/2=1/2+3/k<11/20$, we can pack $v'_1$ in a bin, $v'_2$, $v'_3$, and $v'_5,\ldots,v'_m$ (if they exist) in a second bin, and $v'_4$ in a half bin.
\item[]If $1/2<y_{4,4}<13/20+8/k$, then each of $y_{1,1}\leq1/2$, $y_{2,2}\leq1/2$, and $y_{3,3}\leq1/2$.
If $x_{1,1}\leq1/2$ then we can pack $v'_1$ and in a half bin, $v'_2$ and $v'_3$ in a bin, and $v'_4$ and $v'_5,\ldots,v'_m$ (if they exist) in a second bin. If $x_{1,1}>1/2$ then, both $x_{2,2}\leq1/2$ and $x_{3,3}\leq1/2$, so we can pack $v'_1$ and $v'_3$ in a bin, $v'_2$ in a half bin, and $v'_4$ and $v'_5,\ldots,v'_m$ (if they exist) in a second bin.
\end{description}
\end{description}

\item[]finally, if $x_{a,m}>1$, 
then since $x_{a+1,m}\leq1/20$, the $L_1$-norm $\|v'_a\|_1\geq x_{a,a}>19/20$, so each vector $v'_1,\ldots,v'_{a-1}$ must also have $L_1$-norm greater than $19/20$, thus $a=2$ in this case, as the maximum sum of $L_1$-norms of vectors in a bin is~2. Since $x_{2,2}>19/20$, the value of $x_{1,1}=x_{1,2}-x_{2,2}<1+2/k-19/20=1/20+2/k$, whereby $y_{1,1}=\|v'_1\|_1 -x_{1,1}>19/20-1/20-2/k=9/10-2/k$. Hence, $y_{3,m}=y_{1,m}-y_{2,2}-y_{1,1}<21/20-0-9/10+2/k=3/20+2/k<1/2$. We can therefore pack $v'_1$ in one bin, $v'_2$ in a second bin and $v'_3,\ldots,v'_m$ in a half bin.
\vspace*{-1ex}
\end{description}
This completes the case analysis and proves our lemma.
\hfill$\Box$
\end{proof}

We can now prove the main theorem of this section.
\begin{theorem}\label{thm:log2.5}
The strategy $A_k$ receives $O(\log n)$ bits of advice, works in polynomial time, and has cost\vspace*{-1.5ex}
$$
\frac{5}{2}\big|\OPT(\sigma)\big| + 1,%\vspace*{-0.5ex}
$$
for serving any sequence $\sigma$ of size~$n$, if $k\geq640$ is an even constant.
\end{theorem}
\begin{proof}
As in the proof of Theorem~\ref{thm:limited-log2}, assume first that $A_k$ uses $|A_k(\sigma)|\!=|\OPT(\sdk{\sigma_L})|$ bins, i.e., there is no bin that only contains short vectors. Consider an optimal solution $\OPT(\sigma)$ and remove all the short vectors from the bins in this solution. This is still a feasible solution for the remaining long vectors, thus $|\OPT(\sigma_L)|\leq|\OPT(\sigma)|$. By Lemma~\ref{lem:pack2.5}, we therefore get%\vspace*{-1.5ex}%
\begin{align}
\big|A_k(\sigma)\big| 
& =
\big|\OPT\big(\sdk{\sigma_L}\big)\big|
\leq 
\frac{5}{2}
\big|\OPT(\sigma_L)\big|+1
\leq
\frac{5}{2}
\big|\OPT(\sigma)\big|+1.
\label{eqn:long2.5}
\end{align}

%\vspace*{-1.5ex}%
Assume now that $A_k$ constructs a solution $A_k(\sigma)$ having at least one bin with only short vectors in it. Each of these short vectors did not fit in any of the critical bins originally opened, thus the virtual load of each critical bin is greater than $1-80/k$, ($80/k$ is the maximum $L_1$-norm of a short vector). Since a bin can contain at most $k/20$ long vectors, and each long vector is scaled at most $1/k$ in the $x$-direction and at most $1/k$ in the $y$-direction, the actual load is greater than $1-k/20\cdot(1/k+1/k)-80/k=9/10-80/k$. Since the maximum load of a bin is~$2$, we have in this case, c.f.~Inequalities~(\ref{eqn:loadbound}) and~(\ref{eqn:binbound}),%\vspace*{-1.25ex}%
\begin{align}
\big|A_k(\sigma)\big| 
& \leq 
\left\lceil\frac{2\big|\OPT(\sigma_L)\big|}{9/10-80/k}\right\rceil
\leq
\left(\frac{20}{9}+\frac{1600}{9k}\right)\big|\OPT(\sigma_L)\big|+1
\leq
\frac{5}{2}
\big|\OPT(\sigma_L)\big|+1
\nonumber\\&
\leq
\frac{5}{2}
\big|\OPT(\sigma)\big|+1,
\label{eqn:short2.2}
\end{align}%\vspace*{-3.0ex}\\%
by choosing $k\geq640$ and even. The strategy reads at most %$k^2\lceil\log(\max_{i,j}\{n_{i,j}\}+1)\rceil\leq k^2\lceil\log(|\sigma_L|+1)\rceil\leq
$k^2\lceil\log(n+1)\rceil\in O(\log n)$ bits of advice, since $k$ is constant.

It remains to prove that the solution $\OPT\big(\sdk{\sigma_L}\big)$ can be computed in polynomial time in $|\sigma_L|$. Let \TT\ be the set of different possible bin types using $k$-scaled long vectors. Since at most $k/20$ long vectors of $k^2$ different types can be packed in a bin, we can bound the number of different packing types~by%\vspace*{-0.75ex}%
\begin{align}
|\TT|
&
\leq
\sum_{l=1}^{k/20} \binom{k^2+l-1}{l} %{k^2 \choose l}  
\in 
O\big(k\cdot(k^2)^{k/20}\big)
=
O\big(k^{1+k/10}\big),
\end{align}%\vspace*{-1.75ex}\\%
which is constant. We let $t_{i,j}$ be the number of $k$-scaled long vectors in the $(i,j)$-box for the bin type $t\in\TT$. Given the advice information $n_{1,1}, n_{1,2},\ldots,n_{k,k}$, the number of long vectors in each box, we can formulate a recurrence for the optimal packing solution as%\vspace*{-1.0ex}%
\begin{align}
%\lefteqn{P\big(n_{1,1}, n_{1,2},\ldots,n_{k,k}\big) =}\hspace*{12ex}
P\big(n_{1,1}, n_{1,2},\ldots,n_{k,k}\big) &=
%&\nonumber\\
\min_{t\in\TT} \left\{ P\big(n_{1,1}-t_{1,1}, n_{1,2}-t_{1,2},\ldots,n_{k,k}-t_{k,k}\big) \right\} +1,
\end{align}%\vspace*{-2.25ex}\\%
that we can solve in polynomial time with dynamic programming, since both $k$ and $|\TT|$ are constants, albeit large ones.
%\hfill$\Box$
\end{proof}%

%\noindent
%\vspace*{-1ex}%
%When $k$ is even, we can prove that our analysis is asymptotically tight by considering the following instance of vectors, $m$ copies of the vector $(1/2-2\epsilon,\epsilon)$, $m$ copies of the vector $(1/2+\epsilon,\epsilon)$, and $m$ copies of the vector $(\epsilon,1-2\epsilon)$, for sufficiently small $\epsilon<1/3k$. It is clear that an optimal packing of these vectors consists of $m$ bins. The $k$-scaled vectors become $m$ copies of $(1/2,1/k)$, $(1/2+1/k,1/k)$ and $(1/k,1)$, respectively. Only pairs of the first type of vectors fit together in a bin, hence the minimum number of bins required after $k$-scaling is~$\lceil5m/2\rceil\geq5m/2$.
When $k$ is even, our analysis is asymptotically tight. Consider the following instance of vectors: $s$ copies of the vector $(1/2-2\epsilon,\epsilon)$, $s$ copies of the vector $(1/2+\epsilon,\epsilon)$, and $s$ copies of the vector $(\epsilon,1-2\epsilon)$, for sufficiently small $\epsilon<1/3k$. An optimal packing of these vectors uses $s$ bins. The $k$-scaled vectors become $s$ copies of $(1/2,1/k)$, $(1/2+1/k,1/k)$ and $(1/k,1)$, respectively. Only pairs of the first type of vectors fit together in a bin, hence the minimum number of bins required after $k$-scaling is~$\lceil5s/2\rceil\geq5s/2$, giving a lower bound of~$5/2$.

%\vspace{-1.25ex}

\section{Combining the Results}\label{sec:ctr}%
\begin{figure*}
                %\vspace*{-2.0ex}
			    \begin{center} \small
			    \input{\figdirfullplotcover.pdf_t}
			    \parcaption{\capw}{\label{fullplotcover}Plots for worst case competitive ratios for~$\gamma\in[0,\pi/2]$ for our combined strategy (red), {\sc FirstFit} (green), and {\sc AnyFit} lower bound~(brown).}
			    \end{center}
			    %\vspace*{-3.0ex}
			    \end{figure*}%

\noindent
Combining our presented strategies with the result by Angelopoulos~{\em et~al}.~\cite{angelopoulos2018online} that achieves a competitive ratio of $1.47012+\epsilon$ with constant advice \big(actually $O(\log\epsilon^{-1})$ bits\big) for the one dimensional case, so that we use this strategy when $\gamma=0$, strategy $A_{\gamma}$ when $0<\gamma\leq\pi/2 - 2\arctan(7/15)$ and strategy $A_{k}$, for $k\geq640$ and even, when $\gamma>\pi/2 - 2\arctan(7/15)$, we have the following %immediate
corollary.
\begin{corollary}
For any angle $0\leq\gamma\leq\pi/2$ and $\epsilon>0$, the combined strategy described receives $O(\epsilon^{-1}\log n)$ bits of advice and has cost $c(\gamma,\epsilon)|\OPT(\sigma)| + 1$ for serving any sequence $\sigma$ of size~$n$, where%\vspace*{-0.5ex}%
{\small
$$
c(\gamma,\epsilon)=
\left\{
\begin{array}{ll}
1.47012+\epsilon%\dfrac{3}{2}
&
\mbox{for $\gamma=0$,}
\\
\max\left\{2,\:\dfrac{6}{1+3\tan(\pi/4-\gamma/2)}+\epsilon\right\}\hspace*{0.54em} 
&
\mbox{for $0<\gamma\leq\pi/2 - 2\arctan(7/15)$,}
\\
5/2%\dfrac{5}{2}
&
\mbox{for $\pi/2 - 2\arctan(7/15)<\gamma\leq\pi/2$.}
\end{array}
\right.%\vspace*{-1ex}%
$$
}
\end{corollary}
Figure~\ref{fullplotcover} illustrates the worst case competitive ratio for different values of~$\gamma$.

\section{Conclusions}\label{sec:conc}

We consider the online two-dimensional vector packing problem and show a lower bound of $11/5$ for the competitive ratio of any {\sc AnyFit} strategy. We also show upper bounds spanning between 2 and $5/2$ depending on the angle restrictions placed on the vectors given logarithmic advice, where the currently best competitive strategy has competitive ratio~$27/10$, albeit without using advice.

Interesting open problems include generalizing the lower bound on the competitive ratio to hold for any strategy (without advice) and relating the advice complexity to the competitive ratio, either by giving specific lower bounds on the advice complexity for a given competitive ratio or through some function that relates one with the other.

{\small
%\vspace*{-2ex}%

}

\end{document}